\documentclass[a4paper,twocolumn,fontsize=10pt,DIV=16,abstract=true]{scrartcl}
\pdfoutput=1

\usepackage[utf8]{inputenc}
\usepackage[T1]{fontenc}
\usepackage{lmodern}
\usepackage[protrusion=true,expansion=true]{microtype}

\usepackage[style=ieee, backend=biber]{biblatex}

\usepackage{tikz}
\usepackage{cleveref}
\usepackage{balance}
\usepackage{pifont} % checkmarks
\newcommand{\xmark}{--}%{\ding{55}}

\usepackage{array}
\usepackage{booktabs} % lines
\usepackage{multirow} % multirow formatting
\usepackage{listings}
\def\inline{\lstinline[basicstyle=\ttfamily\small]}

\definecolor{KITblack15}{cmyk}{0,0,0,.15} %  15% black

\begin{document}

\title{Attacks on Dynamic Protocol Detection of Open Source Network Security Monitoring Tools}

\date{}

\author{
Jan Grashöfer\\
\small Karlsruhe Institute of Technology\\
\small Institute of Telematics\\
\small jan.grashoefer@kit.edu
\and
Christian Titze\\
\small Karlsruhe Institute of Technology\\
\small Institute of Telematics\\
\small christian.titze@alumni.kit.edu
\and
Hannes Hartenstein\\
\small Karlsruhe Institute of Technology\\
\small Institute of Telematics\\
\small hannes.hartenstein@kit.edu
}

\maketitle

\begin{abstract}
Protocol detection is the process of determining the application layer protocol in the context of network security monitoring,
which requires a timely and precise decision to enable protocol-specific deep packet inspection.
This task has proven to be complex, as isolated characteristics like port numbers are not sufficient to reliably determine the application layer protocol.
Hence, more dynamic detection approaches have been developed.
In this paper, we analyze the Dynamic Protocol Detection mechanisms employed by popular and widespread open-source network monitoring tools.
We show on the example of HTTP that all analyzed detection mechanisms are vulnerable to evasion attacks, which pose a serious threat to real-world monitoring operations.
We find that the underlying fundamental problem of protocol disambiguation is not adequately addressed in two of three monitoring systems that we analyzed.
To enable adequate operational decisions, this paper highlights the inherent trade-offs within Dynamic Protocol Detection.
\end{abstract}

\section{Introduction}

Common Network Security Monitoring and Intrusion Detection Systems make use of Deep Packet Inspection (DPI) techniques to allow application layer specific analysis of the monitored traffic.
Once the application layer protocol is identified, these systems attach appropriate decoders to extract detailed meta data about or contents of the communication for further analysis.
While the primary focus of a Network Security Monitoring (NSM) system is to provide as detailed information about the observed traffic as possible, access to the high-level semantics of the traffic is extremely valuable for Intrusion Detection Systems as well.
For example, in case of signature-based intrusion detection, the additional information can be used to improve the accuracy of signatures to reduce false positives.
In the following, we will refer to NSM systems in the sense of a superclass, which includes network Intrusion Detection and Prevention Systems (IDS/IPS).

Determining the correct application layer protocol decoder for a connection based on port numbers has proven insufficient.
On the one hand, there are arbitrary deviations from using standardized ports for multiple reasons ranging from web interfaces operated on peculiar ports
to users or applications actively trying to bypass port-based restrictions.
On the other hand, there are protocols that use unpredictable ports by design, because ports are automatically negotiated.
Popular examples are FTP and SIP.
Hence, the concept of Dynamic Protocol Detection (DPD) has evolved,
which denotes a flexible approach that takes the actual content of a connection into account to determine the protocol in use, i.e.\ to perform protocol disambiguation.
DPD has been introduced as a key element to virtually almost all modern NSM systems, because failing to detect the protocol in use prevents the appropriate decoding of traffic and hence spoils the intended visibility.

In this paper, to attack the monitoring, we revisit DPD and transfer general evasion strategies to DPD.
By analyzing state of the art DPD implementations, we deduce two attack techniques, \emph{Deferred Start} and \emph{Misleading Start}, which exploit the underlying problem of protocol disambiguation.
Applying the DPD attack techniques for text-based protocols on the example of HTTP, we implement three practical attack realizations.
Although most of the traffic is encrypted nowadays, HTTP still represents a highly relevant use case:
Apart from scenarios in which the lack of encryption constitutes the reason for careful monitoring, the presented attacks would also affect monitoring operations that deliberately decrypt HTTPS traffic for monitoring purposes \cite{durumeric_security_2017, waked_intercept_2018}.
We show that the attacks pose a threat to real-world deployments by an evaluation of the behavior of popular web servers, such as nginx, when facing our attack traffic.
We find that the web server behavior exploited for the proposed evasion is quite common in the top 500 websites.
Based on the outlined attacks, we point out the inherent trade-offs within DPD and discuss approaches to address them.
Given the fundamental nature of the underlying problem, we intend to raise the awareness for the need to carefully balance these trade-offs.

This paper is structured as follows:
First, we present related work in \ref{sec:related_work}.
In \ref{sec:threat_model} we define the attack scenario and introduce our attack strategy.
Then we analyze the DPD mechanisms employed by two popular and widespread open-source NSM tools, namely Bro%
\footnote{In October 2018 Bro was renamed Zeek. As our analysis focused on Bro version 2.5.1, we use the old name for the remainder of this paper.}
and Snort in \ref{sec:mechanisms}.
Building upon our insights, in \Cref{sec:attack_techniques}, we introduce two general DPD attack techniques that are tailored to our attack scenario and
construct different types of attack traffic to conduct evasion attacks.
In \ref{sec:evaluation} we show that both analyzed NSM systems as well as a third, popular IDS, Suricata, are vulnerable to these evasion attacks.
In this context, we also discovered a Denial of Service (DoS) attack against Bro that has since been reported and fixed.
Furthermore, we evaluate the real-world applicability and impact of the presented attacks.
Considering our findings, we discuss the challenges of DPD in \ref{sec:discussion}.
Finally, we conclude our paper including an outlook on future work in \ref{sec:conclusion}.

\section{Related Work}
\label{sec:related_work}

Attacks on network monitoring in general are a well-known problem \cite{ptacek_insertion_1998, kreibich_network_2001, cheng_evasion_2012}.
Already in \citeyear{ptacek_insertion_1998}, \citeauthor{ptacek_insertion_1998} \cite{ptacek_insertion_1998} introduced a classification of monitoring attacks,
described corresponding attacks on IP as well as TCP level and evaluated monitoring software against these attacks.
\citeauthor*{ptacek_insertion_1998} showed that all tested systems were vulnerable
and came to the conclusion that substantial efforts have to be undertaken to address the shortcomings they discovered.
Nevertheless, more than a decade later \citeauthor*{cheng_evasion_2012} \cite{cheng_evasion_2012} still found popular NSM software vulnerable to well-known attacks.
\citeauthor{roelker_http_2003} \cite{roelker_http_2003} described techniques to evade detection in context of HTTP focusing on exotic encoding to prevent a comprehensive analysis.
Despite previous work in the NSM domain, our work demonstrates that already gathered insights have not been transferred thoroughly to DPD,
which leaves state of the art monitoring software vulnerable to attacks.

Traffic classification is a field related to DPD.
Numerous approaches have been suggested to infer the type of traffic \cite{nguyen_survey_2008} or the application in use \cite{bujlow_independent_2015}.
Machine learning techniques have even been used to classify encrypted traffic \cite{rezaei_deep_2019}.
However, there is a fundamental difference between traffic classification and protocol detection for NSM:
Protocol detection requires a timely and precise decision to enable protocol-specific DPI,
whereas traffic classification allows for a much higher degree of fuzziness,
as it usually aims at providing a rather coarse-grained overview of traffic composition.

To the best of our knowledge, the first scientific work on DPD in context of NSM is the work of \citeauthor*{dreger_dynamic_2006} \cite{dreger_dynamic_2006} that was published in \citeyear{dreger_dynamic_2006}.
\citeauthor*{dreger_dynamic_2006} present a tree-based approach for dealing with ambiguities in the detection process.
The authors carve out the fundamental trade-offs in protocol detection and foresee attacks on the monitor in general.
In contrast, we will present attacks on the protocol detection mechanism itself.
We believe that \citeauthor*{dyer_protocol_2013} \cite{dyer_protocol_2013} were the first to systematically investigate attacks on DPD in context of DPI.
However, their threat model is substantially different from the established thread model for NSM:
They assume that both endpoints are controlled by the attacker.
To circumvent censorship, they tunnel traffic by mimicking benign protocols.
Our work employs the established threat model for NSM \cite{paxson_bro:_1999, dreger_dynamic_2006}:
We assume that only one of the monitored connection endpoints is controlled by the attacker.
This in turn requires attacks on the monitor to be significantly more advanced.
Nevertheless, we show that DPD mechanisms can be exploited to spoil the visibility defenders seek to gain by deploying NSM systems.

\section{Threat Model \& Attack Strategy}
\label{sec:threat_model}

% Glossar:
% - Attack Strategy = General concept that is applied to conduct an attack
% - Attack Class = Group of attacks following the same approach or sharing the same consequences
% - Attack Scenario = Definition of the specifics of the attacked system and its environment
% - Attack Surface = Parts (e.g. input handling) of the system exposed to attacks
% - Attack Vector = Possible path to follow to execute an attack

In this section, we define the attack scenario that constitutes our threat model (\ref{sec:threat_model:scenario}) and present our attack strategy (\ref{sec:threat_model:strategy}).

\begin{figure}[tb]
\centering
\includegraphics[width=\columnwidth]{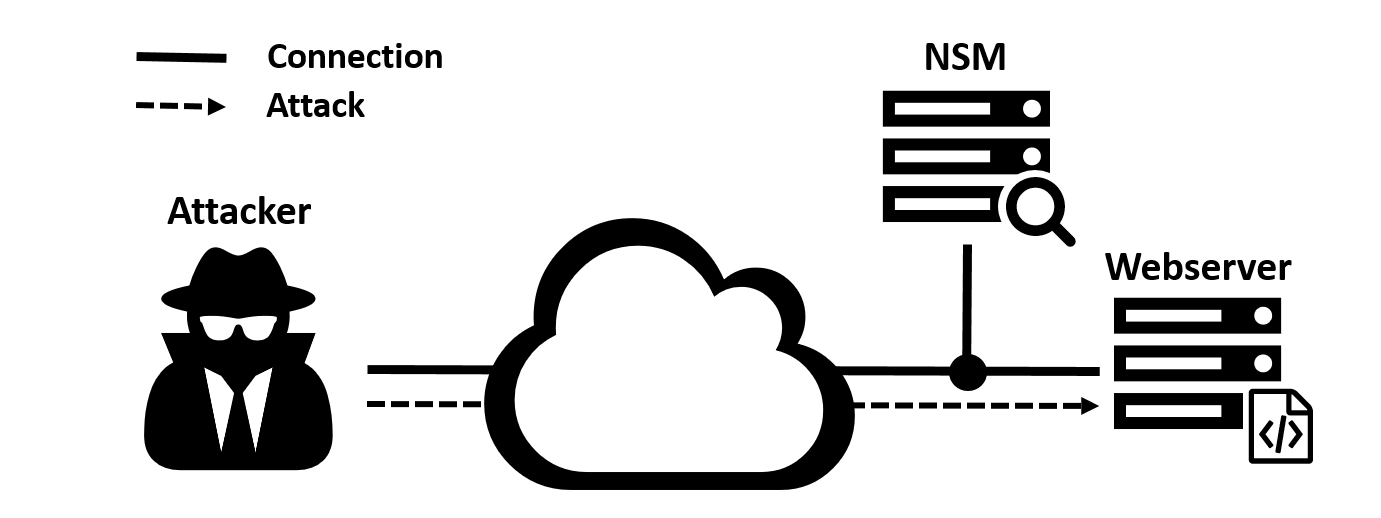}
\caption{
  Attack scenario: The attacker aims at attacking a monitored web server, while hiding the attack traffic from the NSM.
  }
\label{fig:threat_model}
\end{figure}

\subsection{Attack Scenario}
\label{sec:threat_model:scenario}

In this work, we focus on text-based protocols using the example of HTTP.
The attack scenario is depicted in \Cref{fig:threat_model}:
The objective of the attacker is to execute an attack on a web server without being accurately observed by the NSM that monitors the traffic of the web server.
The fundamental assumption in the field of NSM is that only a single connection endpoint is controlled by the attacker.
If both endpoints were under the control of an attacker, the attacker would be able to establish arbitrary covert channels, which in turn are impossible to analyze by an observer \cite{paxson_bro:_1999}.
In our scenario the attacker is able to send arbitrary traffic to the web server, but has no additional means to influence the server's behavior,
i.e. the web server does not cooperate in hiding the attack traffic.
As the NSM receives a full copy of the traffic, the overall attack is twofold:
In addition to the primary objective of attacking the web server,
the attacker also needs to attack the monitor.
Therefor the attacker is required to craft the attack traffic so that it is processed by the web server but evades the NSM.

\subsection{Attack Strategy}
\label{sec:threat_model:strategy}

Our strategy to evade monitoring focuses on the detection of text-based application layer protocols like HTTP, FTP or SMTP.
If we are able to prevent the NSM to correctly detect the protocol in use,
we take away the ability to adequately analyze the observed traffic and thus spoil the visibility that operators of the NSM seek to gain.

While text-based protocols allow for easy analysis by a human observer, parsing these kinds of protocols often represents a difficult task.
Whereas binary protocols encode the length of Application Data Units (ADUs), either in the protocol's definition or in form of a length field, text-based protocol parsers need to accumulate the stream of characters until the end of an instruction, i.e. a line ending, is identified.
The accumulated character string can then be parsed and interpreted according to the protocol.
Unfortunately, it is common for text-based protocols not to limit the total length of ADUs%\cite{langsec}
, which theoretically requires unlimited buffers for accumulating a single ADU.
However, a NSM has to make a timely decision on the protocol.
The less information is available to decide on the protocol in use, the likelier a decision cannot be definitive due to ambiguities.

Application Data Units of text-based protocols are typically represented as distinct lines.
Lines are usually separated using either a single linefeed symbol (LF) or a combination of carriage-return (CR) and linefeed \cite{forshaw_attacking_2017}.
To evade the monitor, we will exploit this general structure of text-based protocols in the context of DPD.
Based on our analysis of protocol detection mechanisms employed in popular NSM systems in the following section,
we will deduce two attack techniques, \emph{Deferred Start} and \emph{Misleading Start}.
These techniques will allow us to artificially delay the point at which a proper decision can be made or deliberately mislead the monitor to make a wrong decision.

Our example focuses on HTTP, due to its prevalence in the modern Internet.
While all of the analyzed DPD mechanisms can be extended to work on HTTP/2 in its unencrypted form%
\footnote{Note that the preface of a HTTP/2 connection is text-based. \cite{belshe_hypertext_2015}},
only one of the investigated monitoring applications comes with an HTTP/2 analyzer.
Thus, we narrowed our analysis down to HTTP/1.1 and below.
Nevertheless, the techniques we apply can also be used for other protocols.

\section{Protocol Detection Mechanisms}
\label{sec:mechanisms}

In this section we will describe the Dynamic Protocol Detection mechanisms of two different open-source NSM systems,
namely Bro \cite{bro_project} and Snort 3 \cite{cisco_snort3}.
To detect the protocol in use independently of the port, both systems apply protocol-specific signatures.
Yet, we will see that the underlying architectures significantly differ.
Note that we omit in this section a detailed analysis of Suricata due to its similarity to Snort.
Furthermore, libraries like nDPI \cite{deri_ndpi_2014} and libprotoident \cite{alcock_libprotoident_2012} do not implement mechanisms to enable DPI based on their protocol classification.
As our work focuses on the inherent trade-offs of such mechanisms, these libraries are out of scope.

\subsection{Bro}
\label{sec:mechanisms:bro}

The Bro NSM serves as a flexible platform for the analysis of network traffic.
Incoming packets are processed by an \emph{event engine} that utilizes protocol analyzers to parse the traffic and generate a high-level event stream.

\begin{figure}[tb]
\centering
\includegraphics[height=1in]{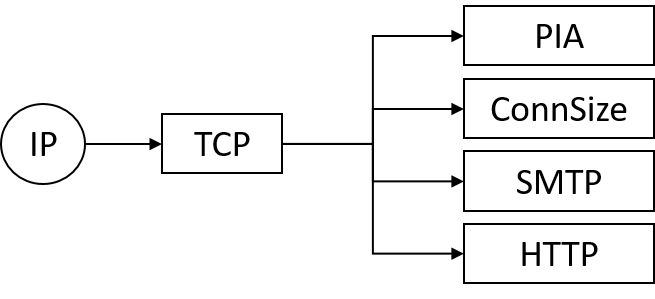}
\caption{
  Exemplary analyzer tree for an IP-based TCP connection in Bro.
  Branches indicate parallel execution of analyzers.
  }
\label{fig:bro_analyzer_tree}
\end{figure}

To detect which protocols are used in the observed network traffic, Bro uses the DPD framework introduced in \citeyear{dreger_dynamic_2006} by \citeauthor*{dreger_dynamic_2006} \cite{dreger_dynamic_2006}.
It orchestrates inspection of a connection by composing a pipeline of protocol analyzers to which the given stream of traffic is fed.
Analyzers can be chained to account for protocols nested into each other.
As it is not always clear which analyzer has to be attached, a protocol analyzer may have multiple child analyzers.
This results in each connection managing its analyzers in the form of a tree as shown in \Cref{fig:bro_analyzer_tree}.
If an analyzer signals a serious inconsistency with its protocol, the analyzer is removed from the tree.
Following this trial-and-error approach, possible ambiguities in the protocol detection can be solved eventually.

\Cref{fig:bro_analyzer_tree} shows an exemplary analyzer tree for a TCP connection.
In this case the \emph{TCP Analyzer} is set as root analyzer and takes care of reassembling the stream.
By default, the \emph{Connection Size Analyzer (ConnSize)}, which keeps track of the connection's statistics like bytes sent and received,
and the so called \emph{Protocol Identification Analyzer (PIA)} are added as child analyzers.
Furthermore, the initial analyzer tree is built based on well-known ports (e.g. $80$ for HTTP).
Besides well-known ports, each protocol analyzer might specify signatures to get triggered.
For example, the SMTP analyzer in \Cref{fig:bro_analyzer_tree} might have been attached due to a signature match.
\Cref{fig:bro_config} shows a pair of signatures for HTTP:
The client signature specifies a set of common HTTP methods and the server signature matches the generic beginning of an HTTP response.
The HTTP analyzer is triggered by the server signature,
which requires the client signature to have matched by specifying the \inline{requires-reverse-signature} condition.
The PIA is responsible for matching the DPD signatures and attaching new analyzers to the tree.
To allow the activation of additional analyzers in the course of a connection, the PIA buffers the beginning of the connection.
The buffer size is configurable and by default set to $1024$ bytes.
If an additional analyzer is added to the tree, the buffer is replayed to that analyzer.
Once the buffer is filled, the dynamic detection process is stopped.

\begin{figure}[tb]
\begin{center}
\begin{lstlisting}
signature dpd_http_client {
  ip-proto == tcp
  payload /^[[:space:]]*(OPTIONS|GET|HEAD|
                         POST|...)[[:space:]]*/
  tcp-state originator
}

signature dpd_http_server {
  ip-proto == tcp
  payload /^HTTP\/[0-9]/
  tcp-state responder
  requires-reverse-signature dpd_http_client
  enable "http"
}
\end{lstlisting}
\end{center}
\caption{\label{fig:bro_config} Excerpt from Bro's signature file that defines a pair of patterns to detect the HTTP protocol.}
\end{figure}

\subsection{Snort 3}
\label{sec:mechanisms:snort}
Snort is likely today's most popular open source IDS and IPS solution and mainly focuses on matching patterns in network traffic.
The patterns are defined in the form of rather low-level rules, matching on the binary stream of network traffic.
Snort 3 allows the restriction of rules to certain application layer protocols like HTTP,
whereas previous versions only supported TCP, UDP, ICMP and IP \cite{roesch_snort_2019}.
In August 2018, a beta version of Snort 3 was released \cite{cisco_snort_blog_2018}
but it has not been declared production ready, yet.
Nevertheless, in context of this paper, we will consider Snort in version 3.

In Snort 3 application layer analysis is realized by so called service inspectors.
A service inspector consists of a stream splitter that preprocesses the data stream and splits it into protocol-specific ADUs, e.g. lines in case of HTTP.
The ADUs are subsequently processed by the actual service inspector, which also manages the protocol state.

\begin{figure}[tb]
\begin{center}
\begin{lstlisting}
http_methods = { 'GET', 'HEAD', 'POST', ... }

default_wizard = {
  spells = {
    { service = 'http', proto = 'tcp',
      client_first = true,
      to_server = http_methods,
      to_client = { 'HTTP/' } },

    { service = 'smtp', proto = 'tcp',
      client_first = true,
      to_server = { 'HELO', 'EHLO' },
      to_client = { '220*SMTP', '220*MAIL' } }, ...
  }, ...
}
\end{lstlisting}
\end{center}
\caption{\label{fig:snort_config} Excerpt from \lstinline[basicstyle=\ttfamily\footnotesize]{snort_defaults.lua} (default configuration) that defines patterns called spells to detect text-based protocols. }
\end{figure}

DPD is implemented by a special service inspector.
To identify text-based protocols, patterns called spells are matched.
Spells are realized as acceptors, i.e. finite state machines that accept a regular language.
\Cref{fig:snort_config} shows the definition of the patterns for HTTP and SMTP.
Alongside the service identifier (\inline{service}), which is used for assigning the corresponding inspector,
each spell defines the patterns expected in a first message to the server (\inline{to_server}) and in its reply (\inline{to_client}).
While the development notes and code architecture suggest that the final decision on the application layer protocol should be on the service inspector that is triggered,
as of writing this paper, Snort 3 relies on the DPD done by the wizard inspector.
Once the wizard attached a service inspector, the classification of the connection is fixed.

Summing up, we find three important facets of DPD:
First, traffic has to be buffered until a decision on the protocol is made.
Second, signatures are used to infer the protocol in use and third,
the applied heuristics might not be definite so that the resulting ambiguities need to be handled.
Considering a third NSM system for our evaluation in \Cref{sec:evaluation}, we will see that Bro is the only system that actively deals with ambiguities by implementing an analyzer tree.

\section{DPD Attack Techniques}
\label{sec:attack_techniques}

In this section we present two techniques to conduct novel insertion attacks on DPD in context of text-based protocols (\ref{sec:attack_techniques:deferred_start} \& \ref{sec:attack_techniques:misleading_start})
and develop three practical realizations of these techniques (\ref{sec:attack_techniques:realization}).
Based on the practical realizations,
we will evaluate the DPD mechanisms implemented in popular open source NSM systems in \Cref{sec:evaluation}.

\subsection{Deferred Start}
\label{sec:attack_techniques:deferred_start}
A \emph{Deferred Start} attack is based on the observation that NSM systems are forced to focus on the beginning of a connection to identify the protocol.
The basic idea is to defer the actual start of the connection by flooding the monitor with useless data,
causing the protocol detection to fail.

Considering that every byte of the connection is relevant for analysis,
the monitored traffic has to be buffered for analysis until the protocol is finally determined.
This requirement is further aggravated by the fact that the beginning of a connection usually contains information of particular interest:
For example, request and response headers in the beginning of an HTTP connection provide meaningful meta data.
As extensive buffering would pose a Denial-of-Service vector, the buffer size has to be limited,
which requires the DPD mechanism to come to a timely decision.

In general there are two possible approaches to deal with the need to buffer traffic in the light of delayed protocol determination:
First, one could use a reasonable sized \textbf{fixed buffer}.
The size has to be chosen large enough to make sure that the protocol can be determined but as small as possible to save valuable resources.
If the protocol cannot be detected based on the buffered traffic,
the DPD mechanism would give up and report a detection failure.
Considering an attacker who might be able to flood that buffer this approach seems suboptimal,
but in fact, this technique implements a trade-off:
For a protocol that is unknown to the monitoring software, the DPD mechanism will never be able to detect the correct protocol.
In this case, continuously trying to detect the protocol would again waste valuable resources.
Thus a fixed buffer implicitly realizes a protocol detection timeout.

Second, a \textbf{ring buffer} could be used.
In contrast to a fixed buffer, a ring buffer of capacity $n$ always provides access to the last $n$ bytes of the connection under consideration.
This sliding window approach would be able to mitigate a deferred start as described above.
In theory, a ring buffer also allows ongoing protocol detection as there is no implicit limit that would stop the process.
Although continuous detection might be undesirable, due to binding resources,
the ring buffer approach is more flexible as it allows to decouple the buffering of connection data and the protocol detection timeout.

\subsection{Misleading Start}
\label{sec:attack_techniques:misleading_start}
A \emph{Misleading Start} attack exploits the focus on the connection start as well.
To detect a protocol on a non-standard port, signatures are employed for identification of known protocols.
For example, a signature for HTTP might match version strings like \inline{HTTP/1.1} to trigger an HTTP-related analysis.
Signatures can be further divided into unidirectional and bidirectional signatures.
Whereas unidirectional signatures match on flows independently, i.e. communication from client to server as well as from server to client,
bidirectional signatures match on a combination of patterns across both directions.

When using signatures to detect protocols, there is an inherent trade-off with respect to the strictness of the signatures.
If a signature is too strict, it will fail to correctly flag all connections that use a given protocol.
If a signature is too loose, ambiguities will arise that have to be resolved to come to a final decision.
Both situations might be exploited by an attacker:
Strict signatures can be evaded by leveraging edge-cases that are accepted by liberal endpoint implementations but not covered by the signature.
Loose signatures can be misused to cause additional load on the monitor either by tricking the software into a resource-intensive inspection of actually non-conforming traffic,
or by complicating the process of solving intentionally induced ambiguities.

For example, an HTTP request starts by specifying the method.
The HTTP method has to be assumed as arbitrary string given the protocol's extensibility \cite{fielding_hypertext_2014}.
Hence, after analyzing the first token of a line there can be multiple options for the underlying protocol.

\subsection{Practical Realization}
\label{sec:attack_techniques:realization}

Based on the previously presented DPD attack techniques, we have deduced three approaches to realize practical attacks:\\
\textbf{CRLF Stuffing}
To realize a \emph{Deferred Start}, we prepend a valid HTTP request with whitespace characters to deliberately delay the protocol detection.
We have chosen the combination of CR and LF characters as they serve as line delimiter for HTTP and thus also occur in valid requests \cite{fielding_hypertext_2014-1}.\\
\textbf{Unknown Method}
The simplest possible approach to realize a \emph{Misleading Start} attack is to use an HTTP method unknown to the monitoring software.
We implement this approach by sending a request that uses the string \lstinline[basicstyle=\ttfamily\footnotesize]{UNKNOWNMETHOD} as method.\\
\textbf{HELO Method}
A more advanced option for a \emph{Misleading Start} is to exploit potential ambiguities that are for example caused by overlapping signatures.
We implement this approach by using the string \inline{HELO} as method,
which is also the first command in an SMTP session.

The last two attacks make use of methods that are not supported by the web server to avoid detection by the NSM system.
We added a \inline{Connection: keep-alive} to our first, manipulated request and send a second, valid request after the first one.
By reusing the same connection we intend to hide the valid, potentially malicious request from the monitor.
The attack is successful if the second request is not detected by the NSM system but correctly processed by the web server.

\section{Evaluation}
\label{sec:evaluation}

In this section, we assess the real-world consequences of the developed attacks with respect to the effect on the monitoring software (\ref{sec:evaluation:impact_on_monitoring}).
For our tests, we use Bro, Snort and Suricata, a third popular open-source NSM system.
Furthermore, we evaluate the susceptibility of web servers to determine the effectiveness of our attacks under real-world conditions (\ref{sec:evaluation:susceptibility}).
Finally we will consider the combined effectiveness (\ref{sec:evaluation:combined}), as our threat-model requires a twofold attack.

\subsection{Attacking the Monitor}
\label{sec:evaluation:impact_on_monitoring}

Our primary interest is the detection behavior for traffic on non-standard ports, as we focus on the dynamic detection of protocols.
Nevertheless, the generated traffic might also impact protocol detection on standard ports.
Hence we conducted our experiments utilizing both a standard port for HTTP ($80$) as well as a non-standard port ($4242$).
\Cref{tbl:nsm_attacks} provides an overview of our findings.
Note that evasions for traffic on port $80$ are of particular severity, because traditional port-based heuristics would have covered them.
To infer the impact of the discovered attacks, we also assess the consequences of successful attacks and discuss possible mitigations.

\begin{table*}[tb]
\caption{Dynamic Protocol Detection vulnerabilities of open source NSM software.}
\label{tbl:nsm_attacks}
\centering
\footnotesize
\begin{tabular}{ l r c c c c c c }
\toprule
	\multirow{2}{*}{\textbf{NSM System}} & \textbf{Attack:} &
	\multicolumn{2}{c}{\textbf{CRLF Stuffing}} &
	\multicolumn{2}{c}{\textbf{Unknown Method}} &
	\multicolumn{2}{c}{\textbf{HELO Method}}\\
	\cmidrule(lr){3-4}
	\cmidrule(lr){5-6}
	\cmidrule(lr){7-8}
	& \textbf{Port:} &
	4242 & 80 & 4242 & 80 &	4242 & 80 \\
\midrule
\multicolumn{2}{l}{Bro 2.5.1                 } & Evasion\textsuperscript{1} & DoS & Evasion & \xmark & Evasion & \xmark\\
\multicolumn{2}{l}{Bro 2.5.1 (unidirectional)} & Evasion\textsuperscript{1} & DoS & \xmark & \xmark & \xmark & \xmark\\
\multicolumn{2}{l}{Snort 3                   } & \xmark\textsuperscript{2} & \xmark\textsuperscript{2} & \xmark & \xmark & Evasion & Evasion \\
\multicolumn{2}{l}{Suricata 4.1.2            } & Evasion\textsuperscript{1} & Evasion\textsuperscript{1} & \xmark & \xmark & Evasion & Evasion \\
\bottomrule
\multicolumn{7}{p{10cm}}{\xmark\ stands for \emph{not} vulnerable}\\
\multicolumn{7}{p{10cm}}{\textsuperscript{1} To evade the (configurable) buffer has to be exhausted.}\\
\multicolumn{7}{p{10cm}}{\textsuperscript{2} HTTP Inspector is attached but cannot cope with the traffic.}\\
\end{tabular}
\end{table*}

\paragraph{Bro}
As described in \Cref{sec:mechanisms:bro}, the Bro NSM uses a fixed buffer to realize a cutoff threshold for protocol detection.
Given a default PIA buffer size of $1024$ bytes, correct protocol detection on non-standard ports is susceptible to the \textbf{CRLF Stuffing} approach,
if a request is prefixed by a sufficiently large number of bytes.
The standard detection signature for HTTP is bidirectional, i.e. it encompasses patterns for request and response headers.
This means that the PIA needs to buffer a complete request and at least the beginning of a response to match the signature.
Hence, the number of stuffing characters required to exhaust the buffer also depends on the length of the request.
In theory detection could be avoided using a sufficiently large, valid HTTP request that suppresses the recognition of the response header.
The possibility of buffer exhaustion has been foreseen by the creators of the DPD framework and was legitimated as deliberate design decision \cite{dreger_dynamic_2006}.
To allow the user to balance the resulting trade-off, the buffer size is configurable.

When confronted with the same traffic on a standard port, Bro successfully detects and analyzes the HTTP session.
This is due to the fact that the HTTP analyzer is already added based on the port when building the initial tree of analyzers.
Furthermore, the HTTP analyzer does not consider the consecutive CRLFs as protocol violation but as empty request lines and thus keeps attached to the session.
However, each empty request triggers a so called weird event causing hundreds of events generated per packet.
Because triggering events on a per packet basis is already considered overly expensive \cite{bro_doc},
this allows to perform a Denial of Service attack.
We reported the Denial of Service attack to the project and it was addressed in the Bro 2.5.5 security release by introducing a sampling mechanism for these events
\cite{bro_release_notes}.

Aside from the Deferred Start, Bro's protocol detection can be evaded in its standard configuration by both Misleading Start attacks,
\textbf{Unknown Method} and \textbf{HELO Method}.
To cause the HTTP analyzer to be attached, the request part as well as the reply part of the signature need to match.
Because the request part is based on known methods only, the signature can easily be evaded using unknown methods.
Like the buffer size, the protocol signatures can be configured freely by the user.
There are two possible solutions to relax the signature:
First, the combined, bidirectional signature can be split into two separate, unidirectional signatures.
In \Cref{tbl:nsm_attacks} this variant is listed as ``Bro (unidirectional)''.
Second, the request part can be improved by matching HTTP version information as well.

In case of Bro, evading protocol detection for HTTP traffic prevents the protocol specific analysis.
In particular, there will be no meta data about the session in the \inline{http.log} file and no HTTP specific event will be generated.
However, general meta data about the connection, e.g. size and duration,
is still being gathered and written to the \inline{conn.log} file.
Although Bro is vulnerable to all of the attacks in its standard configuration,
the system can be configured to withstand them.
With respect to the fixed size of the PIA buffer, which causes the vulnerability to Deferred Start attacks,
a ring buffer approach is preferable, as it provides more flexibility to counter this type of attack.

\paragraph{Snort 3}
The Dynamic Protocol Detection mechanism of Snort 3 is not vulnerable to the \textbf{CRLF Stuffing} attack approach.
As described in \Cref{sec:mechanisms:snort}, the detection of text-based protocols is realized by protocol signatures called spells.
The corresponding acceptors explicitly ignore leading Spaces (SP), Tabs (TAB), Carriage Returns (CR) and Line Feeds (LF).
The set of ignored characters is hard-coded, leaving the opportunity to defer the connection start by using other prefix characters.
While this is theoretically possible, \Cref{sec:evaluation:susceptibility} will show that the approach is not exploitable in practice.
Although the DPD mechanism of Snort 3 is not vulnerable to the stuffing attack,
we noticed that the attached HTTP inspector is unable to reliably parse the session.
For example, in case of 512 CRLFs, two responses and no request is found and
in case of 32 CRLFs only one response could be parsed.
This observation likely indicates a reassembling issue, which is, however, out of scope for this paper.
Because spells implement unidirectional protocol signatures, Snort 3 is also not vulnerable to the \textbf{Unknown Method} attack approach.
Although the request in the attack traffic does not trigger the HTTP detection,
the following reply causes a match and thus causes the whole session to be flagged and analyzed as HTTP.

As Snort 3 does not explicitly consider the case of multiple matching protocol signatures,
it is vulnerable to the \textbf{HELO Method} attack approach:
Once an inspector for a connection is selected, the decision cannot be reverted.
This can be used to trick Snort 3 into a wrong classification, causing an evasion.
Because the \inline{HELO} sequence triggers the SMTP inspector while the sequence is not part of the pattern set to match HTTP (see \Cref{fig:snort_config}),
Snort 3 attaches the SMTP inspector and from here on fails to decode the connection properly.
Given that the attached inspector has to cope with non-conforming traffic,
this behavior bears the potential for Denial of Service attacks.

Due to its strong focus on rule matching, the consequences of evading DPD primarily concern this domain:
Snort 3 offers the possibility to refine rules by specifying HTTP as the protocol (see \Cref{sec:mechanisms:snort}) and
match on selected protocol elements like the headers.
Note that the latter also requires the attached HTTP inspector to correctly parse the session,
which turned out to be problematic during CRLF Stuffing attacks.
If a connection is not classified as HTTP and cannot be inspected accordingly,
Snort does not match any HTTP related rule.
To estimate the real-world impact, we have analyzed popular open rule sets with respect to their use of HTTP-specific rules.
\Cref{tbl:rules} shows that these sets contain a significant number of rules that can be evaded by preventing the correct detection of HTTP traffic.
To work around this issue, one can weaken the specification of the affected rules:
The protocol constraints have to be relaxed, e.g. using TCP instead of HTTP,
and references to protocol elements that would have been made available by the inspector have to be converted into more general expressions.
While this allows the rule to match again,
it introduces overhead by increasing the number of rules that have to be considered for matching on lower protocol levels.
Thus, the workaround increases the chance of false positive matches.
All in all, a significant percentage of common rules can be evaded,
while the available workaround comes with a substantial performance and quality degradation.

\begin{table}[tb]
\caption{HTTP related IDS rules, i.e. rules that rely on correct protocol detection.}
\label{tbl:rules}
\centering
\footnotesize
\begin{tabular}{l r r r}
\toprule
	\multicolumn{1}{c}{\textbf{Rule Set}} &
	\multicolumn{1}{c}{\textbf{Rules Total}} &
	\multicolumn{2}{c}{\textbf{HTTP-related}}\\
\midrule
ET Snort Edge open   & 19.673 &  8.530 & 43\%\\
Snort 3 Community    &    829 &    487 & 58\%\\
ET Suricata 4.0 open & 19.328 & 10.654 & 55\%\\
Positive Research    &    317 &     52 & 16\%\\
\bottomrule
\multicolumn{4}{p{7cm}}{All rule sets have been obtained on 12th of March 2019.}
\end{tabular}
\end{table}

\paragraph{Suricata}

A third, popular open-source Network Security Monitoring system is Suricata \cite{oisf_suricata}, which focuses on intrusion detection and is heavily influenced by Snort.
For our experiments, we used Suricata in version 4.1.2 operated under a standard configuration.
Our black-box test revealed that Suricata is vulnerable to the \textbf{CRLF Stuffing} attack.
But, as the attack requires about one hundred thousand CRLFs to evade protocol detection, it does not pose a serious threat in real-world scenarios as described in \Cref{sec:evaluation:susceptibility}.
In the course of our experiments, we tried to mitigate the attack by increasing the reassembly buffers for TCP without success.
Suricata is not vulnerable to the the \textbf{Unknown Method} attack.
However, we have been able to successfully execute the \textbf{HELO Method} attack,
leaving all tested systems vulnerable.

In the case of Suricata, the consequences of evading the Dynamic Protocol Detection are twofold:
First, Suricata is capable of generating an HTTP log that records meta data about the observed HTTP sessions.
Evading the protocol detection mechanism prevents the software from extracting the meta data and thus suppresses logging of the corresponding information.
Second, given that Suricata was developed as an alternative to Snort, it supports similar rule matching functionality.
Like for Snort, a significant number of rules for Suricata are HTTP-specific as can be seen in \Cref{tbl:rules}.
Again, attacking the DPD mechanism allows evading these rules.
While Snort 3 is officially in beta state, Suricata is deployed in productive environments.
Hence, the severity in case of Suricata should be considered even higher.

\subsection{Susceptibility of Web Servers}
\label{sec:evaluation:susceptibility}
The strategies presented to evade the different monitoring solutions will only pose a threat if the used protocol deviations do not affect the intended recipient, which in our case is an HTTP server.
In this section, we review how popular web servers handle the protocol variations we seek to use for monitoring evasion.

\paragraph{Deferred Start}
\label{sec:susceptibility:deferredstart}

Our analysis of Bro's DPD mechanism revealed the possibility of a Denial of Service attack.
As the DoS vector is solely based on the request, it can be exploited independently of the communication endpoint.
Furthermore, it is possible to evade correct protocol classification for connections to non-common ports, if we are able to fill the PIA buffer.
For this attack vector to pose a threat, we need to determine characters that can be prepended to valid HTTP requests without affecting the interpretation by the web server.
To find suitable prefixes, we generated requests preceded by every possible 16 bit permutation, covering two ASCII characters.
We consider a prefix as ignored, if the server replies with status code 200 (OK) to our request.
\Cref{tbl:webserver_characters} lists ignored characters for each tested web server together with the maximal number of repetitions tolerated.
While most of the tested web servers accept any permutation of CR and LF, Apache just ignores the CRLF sequence.
Only lighttpd does not accept any leading character.
With respect to the default buffering capabilities of Bro, the web servers nginx, nodejs and IIS will offer the opportunity for an attack.
The ignored characters can be prepended to a request without affecting its interpretation by the web server, but will exhaust the DPD buffers of the NSM system.

\begin{table}[tb]
\caption{Leading characters ignored by web servers.}
\label{tbl:webserver_characters}
\centering
\footnotesize
\begin{tabular}{l l l r}
\toprule
	\multicolumn{2}{c}{\textbf{Web Server}} &
	\multicolumn{1}{c}{\textbf{Ignored}} &
	\multicolumn{1}{c}{\textbf{Maximal}}\\
	\multicolumn{2}{c}{\textbf{(Version)}} &
	\multicolumn{1}{c}{\textbf{Characters}} &
	\multicolumn{1}{c}{\textbf{Repetitions}}\\
\midrule
Apache   & (2.4.49) & CRLF            & 20\\
nginx    & (1.14.0) & CR, LF          & $>$10m\\
IIS      & (8.5) & TAB, SP, CR, LF & 16.271\\
lighttpd & (1.4.45) & -               & -\\
nodejs   & (8.10.0) & CR, LF          & 81.797\\
\bottomrule
\end{tabular}
\end{table}

\paragraph{Misleading Start}
\label{MisleadingStart}

To evade Snort 3 and Suricata, we need to mislead the wizard inspector to recognize a different protocol.
As there should be no overlapping between HTTP and another text-based protocol, this requires to send invalid HTTP requests that will not be processed by the web server.
However, if the server does not close the connection, we could send further requests in the same connection that would evade proper analysis by the NSM.
Although we cannot send arbitrary data, the first portion of an HTTP request constitutes the method to use and is thus variable.
According to RFC 7230 \cite{fielding_hypertext_2014-1}, web servers should respond with the status codes 501 (Not Implemented) or 405 (Method Not Allowed) if the method is unknown to the server or not allowed for the requested resource, respectively.
\Cref{tbl:webserver_requests} shows that nginx, IIS and the hardened Apache \cite{owasp_scg_2016} keep the connection open, leading to an exploitable situation.
According to a recent survey, these three web server implementations are used to serve about $75\%$ of all websites \cite{netcraft_august_2019}.
Furthermore, we surveyed the web servers hosting the top 500 websites based on the TRANCO ranking \cite{le_pochat_tranco:_2019}.
We found that $28\%$ of the reachable web servers keep connections open, when confronted with an unimplemented method,
which would offer a potential attack surface.
Note that a large request method string might also be used to fill detection buffers as described in \Cref{sec:attack_techniques:deferred_start}.

\begin{table}[bt]
\caption{Web servers confronted with an unimplemented request method*.}
\label{tbl:webserver_requests}
\centering
\footnotesize
\begin{tabular}{l l l l l}
\toprule
	\multicolumn{2}{c}{\textbf{Web Server}} &
	\multicolumn{2}{c}{\multirow{2}{*}{\textbf{Reaction}}} \\
	\multicolumn{2}{c}{\textbf{(Version)}} & \\
\midrule
Apache & (2.4.29) & 501 Not Implemented &$\rightarrow$ closed\\
Apache & (2.4.29\textsuperscript{\dag}) & 403 Forbidden &$\rightarrow$ \textbf{open}\\
nginx & (1.14.0) & 405 Method Not Allowed &$\rightarrow$ \textbf{open}\\
IIS & (8.5) & 405 Method Not Allowed &$\rightarrow$ \textbf{open}\\
lighttpd & (1.4.45) & 501 Not Implemented &$\rightarrow$ closed\\
lighttpd & (1.4.45\textsuperscript{\dag}) & 501 Not Implemented &$\rightarrow$ closed\\
nodejs & (8.10.0) & \multicolumn{2}{c}{closed immediately}\\
\bottomrule
\multicolumn{4}{p{7.5cm}}{\dag\ The web server was set up to use a ``hardened'' configuration, which limits the available request methods.}\\
\multicolumn{4}{p{7.5cm}}{* For testing the unimplemented method behavior the string ``\inline{UNKNOWNMETHOD}'' was used as HTTP method.}
\end{tabular}
\end{table}

\subsection{Combined Effectiveness}
\label{sec:evaluation:combined}

As discussed in \Cref{sec:threat_model:scenario}, the overall attack is twofold.
While all analyzed NSM systems can be evaded, the attacker needs to combine an attack on the target web server with a suitable evasion approach for the NSM system in place.
For example, the \textbf{HELO Method} approach prevents the port-independent protocol detection for all NSMs in their default configuration.
In case of Snort 3 and Suricata even traffic on well-known ports remains undetected.
To allow a follow-up request after the misleading one, the web server is required to keep the connection open in case of encountering an unimplemented method.
In this scenario, nginx, IIS and the hardened Apache server would offer an attack surface.

\section{Discussion}
\label{sec:discussion}

The underlying issue in detecting an application layer protocol is the fact
that a connection's ports serve as a session identifier and at the same time only implicitly codify the type of the session, i.e. the protocol in use.
The endpoints themselves know about the services they operate by definition and only have to verify that the other endpoint adheres to the corresponding protocol.
While the ports can be used by an observer to track a session,
an observer lacks the background information about which services are operated on which ports.
Therefore, a more general solution to the problem would be to make that information available to the monitor.
If it is not available to the monitor, e.g. because the ports are dynamically negotiated,
the protocol in use has to be inferred based on the observed communication.
There are two fundamental challenges that have to be addressed in this context.

First, \textbf{ambiguities} have to be explicitly considered:
The ideal acceptor for a protocol would be its analyzer.
As many protocols are based on high-level grammars, i.e. they form context sensitive (Chomsky 1) or recursively enumerable (Chomsky 0) languages,
the corresponding analyzers exhibit a significant complexity (or even face decidability issues) \cite{sassaman_halting_2011}.
Thus, to decide on the protocol in use, all analyzers would have to be executed in parallel,
which is inapplicable with respect to the resulting performance needs.
Hence, simpler signatures, e.g. based on context-free grammars (Chomsky 2), are used (see \ref{sec:mechanisms}).
While this approach allows for efficient filtering of the possible options,
it is obvious that the protocol identification cannot be decided based on these signatures.
This means that, after matching signatures, there are ambiguities,
which leave multiple possible options for further analysis.
These options have to be considered by design to come to a well-grounded decision.

Second, the \textbf{cost-benefit ratio} of ongoing protocol detection has to be taken into account:
The trade-off between performance and accuracy by either giving up on a protocol as soon as any protocol violation is encountered
or continuous protocol detection throughout the stream was already described by \citeauthor*{dreger_dynamic_2006} \cite{dreger_dynamic_2006} in the context of Bro.
While they also described attacks on the analyzers, attacks on the DPD mechanisms itself have not been considered explicitly.
But, an attacker might try to mislead a protocol detection mechanism,
either by exploiting an ambiguity or by deliberately switching the protocol in the course of the connection.
In the first case, dealing with ambiguities in general also mitigates this attack.
In the second case, the attached analyzer starts to fail without a possibility to resynchronize again.
To address this situation, protocol detection could be restarted when an analysis begins to fail.
This example underlines that buffering and continuous detection are separate aspects:
While the process of detection is potentially ongoing,
buffering is only needed back to the point where the protocol changes.
Thus buffer size and the threshold for protocol detection should be decoupled.
In addition, buffer contents could optionally be stored in persistent storage for later analysis if the protocol detection failed.

Recapitulating the presented attacks, we would also like to emphasize the exemplary nature of the selected scenario.
Given the trend towards \textbf{encryption} of communication, in particular regarding the Web and the Internet,
we expect the majority of HTTP traffic to be encrypted sooner or later, which prevents content analysis at arbitrary points of the communication path.
However, NSM is usually conducted within the scope of a single management domain.
Thus, the use of devices that terminate encrypted connections and allow access to the unencrypted traffic in a trusted environment is common.
Examples range from load balancing scenarios in which encrypted connections are terminated upfront to the deployment of dedicated TLS-proxies that decrypt traffic on the fly \cite{durumeric_security_2017, waked_intercept_2018}.
Consequently, the challenges and attacks presented in this paper also apply to the analysis of decrypted traffic.
Even without access to the plain text of encrypted traffic, DPD retains its relevance:
On the one hand, DPD is relevant in encrypted environments as collecting protocol-specific meta data is still valuable (e.g., for TLS)
and requires the identification of the protocol in use.
On the other hand, apart from the Internet, there are numerous kinds of networks in which monitoring is indispensable because of a lack of encryption.
This can be either due to external constraints, e.g. the requirement to operate legacy systems,
or a deliberate decision in order to provide transparency:
The alternative to monitoring a networked system in a scenario in which the communication cannot be observed anymore is to gather data on the endpoints.
However, this approach is inapplicable if the endpoints cannot be trusted.

Furthermore, the investigation of DPD also brings a new perspective to the discussion of \textbf{Postel's Law}.
Postel's Law, also known as the Robustness Principle, refers to the implementation of protocols and
states that one should strictly adhere to the standards when sending, but be liberal in what to accept when receiving \cite{carpenter_architectural_1996}.
While the benefits of this approach are obvious, numerous discussions have revealed significant issues for the Internet ecosystem caused by following this rule \cite{thomson_harmful_2019, sassaman_patch_2012}.
So far security considerations primarily criticized the increased complexity of liberal implementations,
which in turn increases the potential for introducing bugs.
Our work demonstrates that following the Robustness Principle can also severely impede the observation of a networked system:
The lax handling of protocol divergences and potential ambiguities result in a vast amount of possible interpretations that have to be considered by a monitor.
Facing this problem, \citeauthor*{kreibich_network_2001} proposed to introduce a component that normalizes traffic \cite{kreibich_network_2001}.
This way a consistent, unambiguous standard is enforced, which can be relied on by the monitor.
However, the suggested approach just shifts the attack surface away from the monitor, towards the normalizer.
Overall, following Postel's law introduces room for interpretation that, in the end, allows for misinterpretation.

All in all, our work shows that the complexity of Dynamic Protocol Detection is prevalently underestimated.
Although well-suited approaches exist that allow to deal with this complexity and balance the resulting trade-offs,
we showed that all tested NSM systems are vulnerable to evasion by exploiting their DPD mechanisms.
With respect to the fundamental nature of the underlying problem and the possible consequences for operating NSM systems,
awareness has to be raised:
Developers need to consider the lessons learned in their designs,
to allow practitioners to balance the inevitable trade-offs.

\section{Conclusion}
\label{sec:conclusion}

In this paper, we analyzed the Dynamic Protocol Detection mechanisms employed by two popular and widespread open-source network monitoring tools.
Building upon our insights, we deduced different DPD attack approaches that focus on the example of HTTP.
Confronting three network monitoring tools with the generated traffic, we were able to evade all of them.
In addition, we discovered a DoS attack in one of the systems, which has been reported and is now fixed.
Given the shortcomings of the state of the art DPD mechanisms,
we evaluated the real-world applicability and impact of the presented attacks.
Based on our results, we come to the conclusion that deficient protocol detection can have a serious impact on the monitoring and security operations.
Considering our findings, we discussed the main challenges of DPD for Network Security Monitoring:
Detection decisions are neither clear nor definite given potential ambiguities in the detection process and
various trade-offs have to be carefully taken into account when trading resources like computation time and memory for accuracy.
The detection process can be tuned in terms of strictness of the detection signatures (e.g., uni- or bidirectional), endurance of the analysis (resource-saving cutoff vs. continuous detection) as well as buffer type (ring vs. linear) and size.
For future work, we want to investigate mechanisms that allow resynchronization of partial streams,
e.g. the mechanism employed by \cite{sommer_spicy:_2016}, with respect to their resilience against attacks.

\section*{Ethical Considerations}
Following the concept of responsible disclosure, we have reported the DoS vulnerability we found in Bro during our experiments to the developers.
Accordingly, the issue was fixed and a security release was published.
In contrast to the DoS attack, the possibilities to evade monitoring are based on a fundamental problem.
By publishing our results we hope to raise the awareness of the resulting trade-offs for practitioners, who have to balance them, and developers, who have to provide the means to do so.

\balance
\printbibliography

\end{document}